\documentstyle[graphics,aps,manuscript,floats,tighten]{revtex}

\makeatletter

\providecommand{\LyX}{L\kern-.1667em\lower.25em\hbox{Y}\kern-.125emX\@}

\newcommand{\X}{{(\mu_{\rho}/2)}}
\makeatother

\begin{document}

\preprint{ADP-01-16/T451}

\title{Chiral Behaviour of the Rho Meson in Lattice QCD}

\author{D. B. Leinweber, A.W. Thomas, K. Tsushima and S. V. Wright}

\address{Department of Physics and Mathematical Physics\\
and Special Research Centre for the Subatomic Structure of Matter,\\
University of Adelaide, Adelaide 5005, Australia}

\maketitle
\begin{abstract}
In order to guide the extrapolation of the mass of the rho meson calculated
in lattice QCD with dynamical fermions, we study the contributions to its self-energy
which vary most rapidly as the quark mass approaches zero; from the processes
\( \rho \rightarrow \omega \pi  \) and \( \rho \rightarrow \pi \pi  \). It
turns out that in analysing the most recent data from CP-PACS it is crucial
to estimate the self-energy from \( \rho \rightarrow \pi \pi  \) using the
same grid of discrete momenta as included implicitly in the lattice simulation.
The correction associated with the continuum, infinite volume limit can then
be found by calculating the corresponding integrals exactly. Our error analysis
suggests that a factor of 10 improvement in statistics at the lowest quark mass
for which data currently exists would allow one to determine the physical rho
mass to within 5\%. Finally, our analysis throws new light on a long-standing
problem with the \( J \)-parameter.
\end{abstract}

\section{Introduction}

As the lightest vector meson, the \( \rho  \) is of fundamental importance
in the task of deriving hadron properties from QCD. Within lattice QCD the ratio
of \( \pi  \) to \( \rho  \) masses is often used as a measure of the approach
to the chiral limit. For a long time lattice calculations were restricted to
values of \( m_{\pi }/m_{\rho } \) above 0.8. However, with the remarkable
improvements in actions, algorithms and computing power, there are now lattice
QCD results with dynamical fermions available for hadron masses with current
quark masses such that \( m_{\pi }/m_{\rho } \) is entering the chiral regime.
Nevertheless, in order to compare with the properties of physical hadrons it
is still necessary to extrapolate the results to realistic quark masses \cite{Leinweber:1999ig}. 

In the past few years there have been some very promising developments in our
understanding of how to extrapolate lattice data for hadron properties, such
as mass \cite{Leinweber:1999ig}, magnetic moments \cite{Magnetic-Moments},
charge radii \cite{Hackett-Jones:2000js} and the moments of structure functions
\cite{Detmold:2001jb}, to the physical region. In doing so it is vital to include
the rapid variation at small pion masses associated with those pion loops which
yield the leading and next-to-leading non-analytic behaviour. (This was crucial
in arriving at a reasonable value for the sigma commutator \cite{Leinweber:2000sa},
for example.) However, a formal expansion of hadron properties in terms of \( m_{\pi } \)
fails to converge up to the region where lattice data exists. The crucial physics
insight which renders an accurate chiral extrapolation possible is that the
source of the pion field is a complex system of quarks and gluons, with a finite
size characterised by a scale \( \Lambda  \). When the pion mass is greater
than \( \Lambda  \), so that the Compton wavelength of the pion is smaller
than the extended source, pion loops are suppressed as powers of \( m_{\pi }/\Lambda  \)
and hadron properties are smooth, slowly varying functions of the quark mass.
However, for pion Compton wavelengths bigger than the source (\( m_{\pi }<\Lambda  \))
one sees rapid, non-linear variations. Phenomenologically this transition occurs
at \( m_{\pi }\sim 500 \) MeV, or \( m_{\pi }/m_{\rho } \) around 0.5 -- the
region now being addressed by lattice simulations with dynamical fermions.

Another difficulty associated with the extrapolation of lattice results that
needs further investigation is the discretisation of momenta inherent in all
lattice calculations. In this regard we mention not only the finite lattice
spacing but the fact that there is a minimum possible non-zero momentum available
because of the finite volume of the lattice. This issue is absolutely critical
to the interpretation of the recent CP-PACS data for dynamical fermions \cite{Aoki:1999ff},
in which a first result\footnote{%
Although CP-PACS finds no evidence of residual errors for the lowest mass point,
they caution that it is premature to draw firm conclusions based on the present
low statistics.
} is reported at \( m_{\pi }/m_{\rho }\sim 0.4 \). As we explain in detail,
the only reason that it is possible to measure the \( \rho  \) mass there is
that the calculation is done in a finite volume. We show that taking the finite
lattice size and finite lattice spacing into account is a necessary requirement
when extrapolating to the physical region. These effects become especially significant
for the case of the \( \rho  \) meson which has a \( p \)-wave, two-pion decay
mode.

In the next section we summarise the key physical ideas and the necessary formulas
for extrapolating the mass of the \( \rho  \) meson to the physical pion mass.
This includes a discussion of the limiting behaviour at small and large quark
mass. We then show the result of our fitting procedure and analyse the uncertainty
in extracting the \( \rho  \) mass at the physical point. We show that a factor
of 10 increase in the number of gauge field configurations at the lowest quark
mass presently accessible would be sufficient to determine the physical \( \rho  \)
mass to within 5\%. In section \ref{sec:J Parameter} we discuss the consequences
of this analysis for the \( J \)-parameter and conclude with a brief summary
and outlook for the future.

\section{Chiral Extrapolation Formula}

\begin{figure}[t]
{\par\centering \resizebox*{0.9\textwidth}{!}{\includegraphics{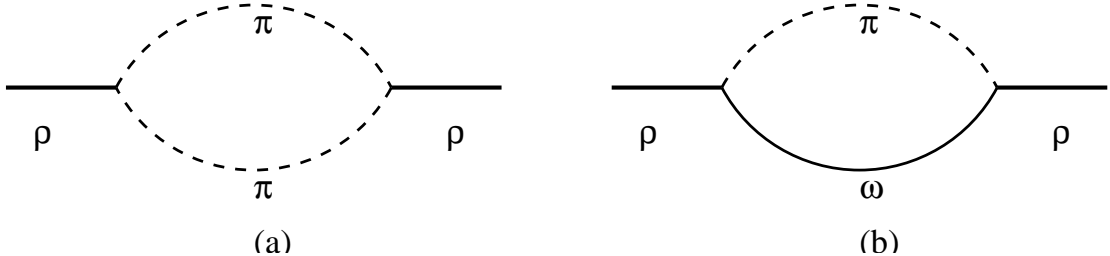}} \par}

\caption{The most significant self-energy contributions to the \protect\( \rho \protect \)
meson mass.\label{fig:SE}}
\end{figure}

The success of our earlier work concerning the extrapolation of the \( N \)
and \( \Delta  \) masses \cite{Leinweber:1999ig} leads us to consider a similar
approach to the latest two-flavour, dynamical QCD data on the \( \rho  \) meson
\cite{Aoki:1999ff,Allton:1998gi}. Once again our guiding principle is to retain
those self-energy contributions which yield the most rapid variation with \( m_{\pi } \)
near the chiral limit -- i.e. those terms which yield the leading non-analytic
(LNA) behaviour and the dominant next-to-leading non-analytic (NLNA) behaviour.
These processes are illustrated in Fig.\ \ref{fig:SE}. The \( \rho \rightarrow \omega \pi  \)
term, shown in Fig.\ \ref{fig:SE}(b), yields the LNA contribution to the \( \rho  \)
mass. The \( \rho \rightarrow \pi \pi  \) term (Fig.\ \ref{fig:SE}(a)) not
only yields the NLNA behaviour but, of course, the width of the \( \rho  \)
once \( m_{\pi } \) goes below \( m_{\rho }/2 \).

In order to evaluate these self-energy terms we take the usual interactions
\cite{Carruthers:1966,Bhaduri:1988gc}:

\begin{equation}
{\cal L}_{\rho \pi \pi }=\frac{1}{2}f_{\rho \pi \pi }\, \vec{\rho }^{\, \mu }\cdot \left( \vec{\pi }\times \left( \partial _{\mu }\vec{\pi }\right) -\left( \partial _{\mu }\vec{\pi }\right) \times \vec{\pi }\right) \, ,
\end{equation}
and 
\begin{equation}
{\cal L}_{\omega \rho \pi }=g_{\omega \rho \pi }\, \varepsilon _{\mu \nu \alpha \beta }\left( \partial ^{\mu }\omega ^{\nu }\right) \left( \partial ^{\alpha }\vec{\rho }^{\, \beta }\right) \cdot \vec{\pi }\, .
\end{equation}
 These lead to the following expressions in the limit where the mass of the
vector mesons (\( \rho  \) and \( \omega  \), taken to be degenerate) is much
bigger than the mass of the pion:

\begin{eqnarray}
\Sigma ^{\rho }_{\pi \pi } & = & -\frac{f^{2}_{\rho \pi \pi }}{6\pi ^{2}}\int _{0}^{\infty }\frac{dk\; k^{4}u_{\pi \pi }^{2}(k)}{w_{\pi }(k)(w^{2}_{\pi }(k)-\mu ^{2}_{\rho }/4)}\, ,\label{eqn:Self Energy pi pi} \\
\Sigma _{\pi \omega }^{\rho } & = & -\frac{g^{2}_{\omega \rho \pi }\mu _{\rho }}{12\pi ^{2}}\int _{0}^{\infty }\frac{dk\; k^{4}u_{\pi \omega }^{2}(k)}{w^{2}_{\pi }(k)}\, .\label{eqn:Self Energy pi omega} 
\end{eqnarray}
In analogy with the heavy baryon limit, we have neglected the kinetic energy
of the heavy vector mesons. Here \( \Sigma ^{\rho }_{\pi \omega } \) and \( \Sigma _{\pi \pi }^{\rho } \)
correspond to the processes shown in Figs.\ \ref{fig:SE}(a), and \ref{fig:SE}(b),
respectively. The pion energy is given by \( w_{\pi }(k)=\sqrt{k^{2}+m_{\pi }^{2}} \),
and \( u_{\pi \pi } \) and \( u_{\pi \omega } \) are dipole form factors governed
by a mass parameter reflecting the finite size of the pion source. In the chiral
limit these have the standard LNA and NLNA behaviour (independent of the forms
chosen for \( u_{\pi \pi } \) and \( u_{\pi \omega } \)) : \textbf{}
\begin{eqnarray}
\left. \Sigma _{\pi \pi }^{\rho }\right| _{{\rm NLNA}} & = & -\frac{f_{\rho \pi \pi }^{2}}{4\pi ^{2}\mu _{\rho }^{2}}m_{\pi }^{4}\ln (m_{\pi })\, ,\nonumber \\
\left. \Sigma _{\pi \omega }^{\rho }\right| _{{\rm LNA}} & = & -\frac{\mu _{\rho }g^{2}_{\omega \rho \pi }}{24\pi }m_{\pi }^{3}\, ,\label{eqn:LNA} 
\end{eqnarray}
 while they are suppressed as inverse powers of \( m_{\pi } \) once \( m_{\pi } \)
is comparable with the dipole mass parameter.\footnote{%
Note that all masses (e.g. the \( \rho  \) mass, \( \mu _{\rho } \)) and coupling
constants should, in principle, be evaluated in the chiral limit. However, as
the variations from the physical values are typically of the order 10\%, we
use the physical values.
} Finally, the \( \rho \rightarrow \pi \pi  \) term contains the unitarity cut
for \( m_{\pi }<\mu _{\rho }/2 \) (as well as an imaginary piece determined
by the width).

The formal solution to the Dyson-Schwinger equation for the \( \rho  \) propagator
places the self-energy contributions in the denominator of the propagator and
thereby modifies the \( \rho  \) mass as \cite{Leinweber:1994yw} :
\begin{eqnarray}
m_{\rho } & = & \sqrt{m_{0}^{2}+\Sigma }\nonumber \\
 & \approx  & m_{0}+\frac{\Sigma }{2m_{0}}\label{eqn:rhoprop} 
\end{eqnarray}
 where \( \Sigma =\Sigma ^{\rho }_{\pi \pi }+\Sigma ^{\rho }_{\pi \omega } \)
and the bare mass, \( m_{0} \), is taken to be analytic in the quark mass.
Guided by the lattice data at large \( m_{\pi } \) we will take \( m_{0} \)
to be \( c_{0}+c_{2}m_{\pi }^{2} \).

\begin{figure}[t]
{\par\centering \resizebox*{0.9\columnwidth}{!}{\rotatebox{90}{\includegraphics{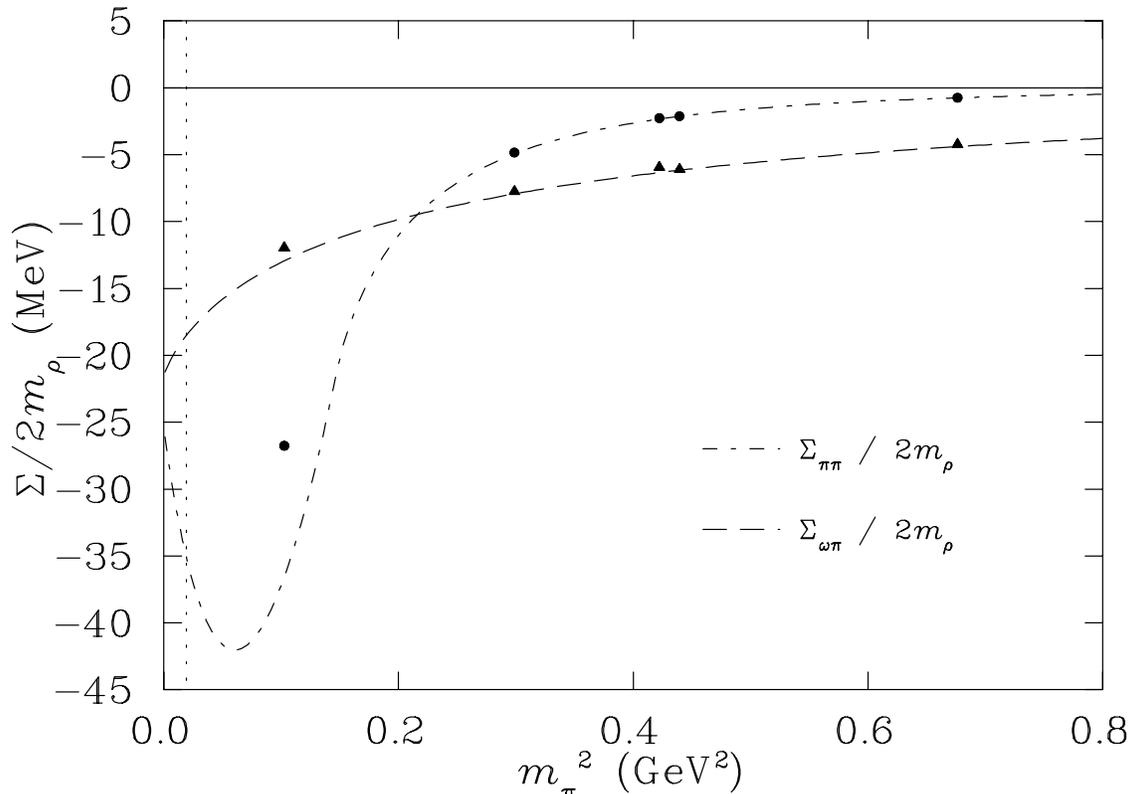}}} \par}

\caption{Variation with pion mass of the self-energy contributions to the \protect\( \rho \protect \)
meson, Eqs.\ (\ref{eqn:Self Energy pi pi}) and (\ref{eqn:Self Energy pi omega}),
for a dipole form factor with \protect\( \Lambda _{\pi \omega }=630\protect \)
MeV . The solid points indicate the value of the self-energy when calculated
at the discrete momenta allowed on the lattices considered in this investigation.
The difference between the curves and points is an indication of the physics
missing because of finite lattice size and spacing.\label{fig:SE diffs}}
\end{figure}
The dipole form factors are defined as

\begin{eqnarray}
u_{\pi \pi }(k) & = & \left( \frac{\Lambda _{\pi \pi }^{2}+\mu _{\rho }^{2}}{\Lambda _{\pi \pi }^{2}+4w_{\pi }^{2}}\right) ^{2}\, ,\label{eqn:Form Factor pi pi} \\
u_{\pi \omega }(k) & = & \left( \frac{\Lambda _{\pi \omega }^{2}-\mu _{\pi }^{2}}{\Lambda _{\pi \omega }^{2}+k^{2}}\right) ^{2}\, ,\label{eqn:Form Factor omega pi} 
\end{eqnarray}
 where \( \mu _{\pi } \) and \( \mu _{\rho } \) are the physical masses of
the \( \pi  \) and \( \rho  \) mesons. The normalisation of \( u_{\pi \pi } \)
is chosen to be unity at the \( \rho  \) pole and the coupling constant, \( f_{\rho \pi \pi }=6.028 \),
is chosen to reproduce the width of the \( \rho  \) (i.e., the imaginary part
of the self-energy). In the \( \rho \rightarrow \omega \pi  \) case we take
\( g_{\omega \rho \pi }=16 \) GeV\( ^{-1} \) \cite{Lublinsky:1997yf}. The
\( m_{\pi }^{2} \) dependence of the self-energies of (\ref{eqn:Self Energy pi pi})
and (\ref{eqn:Self Energy pi omega}) is shown in Fig.\ \ref{fig:SE diffs}
by the dot-dash and dashed curves respectively. The interesting behaviour of
the \( \rho \rightarrow \pi \pi  \) self-energy has been noted in many earlier
works. In the context of lattice QCD it was discussed by DeGrand \cite{DeGrand:1991ip}
and by Leinweber and Cohen \cite{Leinweber:1994yw} and most recently by Szczepaniak
and Swanson\cite{Szczepaniak:2000bi}. Other studies have looked at the self-energy
as a function of \( p^{2} \) (invariant mass of the vector meson) for mixed
\( m_{\pi } \) \cite{Pichowsky:1999mu,Mitchell:1997dn,Hollenberg:1992nj}.

Finally, the lattice data alone cannot separately determine \( \Lambda _{\pi \pi } \)
and \( \Lambda _{\pi \omega } \). In order to constrain them we have therefore
made the reasonable, physical assumption that the size of the source of the
pion field should be the same regardless of whether the intermediate state involves
an \( \omega  \) or a \( \pi  \). Thus we require that \( \Lambda _{\pi \pi } \)
is chosen so as to reproduce the same mean-square radius of the source as would
be generated by the choice of \( \Lambda _{\pi \omega } \). Equating the mean-square
radii results in the following relationship:
\begin{equation}
\Lambda _{\pi \pi }=2\sqrt{\Lambda _{\pi \omega }^{2}-\mu _{\pi }^{2}}\, .
\end{equation}
 An alternative procedure, which could be imposed in future analyses would be
to constrain the difference in the meson self-energy terms to reproduce the
observed \( \rho -\omega  \) mass difference \cite{Pichowsky:1999mu,Mitchell:1997dn,Hollenberg:1992nj,OZI-Rule}.

\subsection{Fitting Procedure}

As we hinted in the introduction, the fact that CP-PACS is able to extract a
measurement of the \( \rho  \) mass at \( m_{\pi }/m_{\rho }<0.5 \) is at
first sight extremely surprising. Once the \( \rho  \) is able to decay one
would expect to measure not the \( \rho  \) mass but the two-pion threshold.
The origin of this result is the quantisation of the pion momentum on the lattice
and in particular the fact that the lowest (non-zero) pion momentum available
is \( 2\pi /aL \) where is \( L \) is the spatial dimension of the lattice.
For the relatively small lattice used by CP-PACS at the lowest pion mass this
corresponds to more than 400 MeV/c momentum. This is why the \( \rho  \) remains
stable.

Motivated by Eq.\ (\ref{eqn:rhoprop}), and wishing to preserve the correct
leading non-analytic behaviour of the self-energies, we have chosen to fit the
\( \rho  \) mass with the simple phenomenological form:
\begin{equation}
\label{eqn:Chiral_Formula}
m_{\rho }=c_{0}+c_{2}m_{\pi }^{2}+\frac{\Sigma ^{\rho }_{\pi \omega }(\Lambda _{\pi \omega },m_{\pi })+\Sigma ^{\rho }_{\pi \pi }(\Lambda _{\pi \pi },m_{\pi })}{2\left( c_{0}+c_{2}m_{\pi }^{2}\right) }\, .
\end{equation}
 Given the constraint relating \( \Lambda _{\pi \pi } \) and \( \Lambda _{\omega \pi } \),
this involves three adjustable parameters. At large \( m_{\pi } \) the self-energies
are suppressed by inverse powers of \( m_{\pi } \) and the \( \rho  \) mass
becomes a simple linear function of \( m_{\pi }^{2} \) (or the quark mass).

In the finite periodic volume, of the lattice, the available momenta, \( k \),
are discrete 
\begin{equation}
\label{eqn:k_mu}
k_{\mu }=\frac{2\pi n_{\mu }}{aL_{\mu }}\, ,
\end{equation}
 where \( L_{\mu } \) is the number of lattice sites in the \( \mu  \) direction,
and the integer \( n_{\mu } \) obeys
\begin{equation}
\label{eqn:n_mu}
-\frac{L_{\mu }}{2}<n_{\mu }\leq \frac{L_{\mu }}{2}\, .
\end{equation}
Therefore to simulate the calculations that are done on the lattice, we should
replace the continuous integrals over \( k \) in Eqs.\ (\ref{eqn:Self Energy pi pi})
and (\ref{eqn:Self Energy pi omega}) with a discrete sum over \( |\vec{k}| \).
However when \( |\vec{k}| \) is zero, the case of a pion emitted with zero
momentum, the integrands vanish, and hence do not contribute to the self-energy.
In fact there is no contribution to the self-energies until \( k_{\mu }=\pm 2\pi /aL_{\mu } \).
There is therefore a momentum gap on the lattice for \( p \)-wave channels,
produced by this discretisation of momenta.

We have investigated this momentum dependence by evaluating the self-energy
integrals, Eqs.\ (\ref{eqn:Self Energy pi pi}) and (\ref{eqn:Self Energy pi omega}),
by summing the integrand at the allowed values of the lattice 3-momenta
\[
4\pi \int ^{\infty }_{0}k^{2}dk=\int d^{3}k\approx \frac{1}{V}\left( \frac{2\pi }{a}\right) ^{3}\sum _{k_{x},k_{y},k_{z}}\, ,\]
 where the \( k_{\mu } \) are defined by Eqs.\ (\ref{eqn:k_mu}) and (\ref{eqn:n_mu})
and \( V \) is the spatial volume of the lattice. The results for the self-energy
contributions are presented in Fig.\ \ref{fig:SE diffs}. The self-energy calculated
on the lattice (the solid circles and triangles) differs little from the full
self-energy calculation in the high quark mass (\( m_{\pi }^{2} \)) region.
Furthermore, the effect in the \( \rho \rightarrow \omega \pi  \) self-energy
contribution is also small at low pion mass. The biggest change is in the \( \rho \rightarrow \pi \pi  \)
self-energy calculation, at lower quark mass. This is the region in which one
might expect the biggest corrections because one is approximating a principal
value integral on a finite mesh. This change in behaviour, particularly at the
lowest data point (\( m_{\pi }^{2}\approx 0.1 \) GeV\( ^{2} \)), indicates
that the \( \pi \pi  \) self-energy contribution is significantly understated
in the lattice simulations. Upon calculating the full self-energy contribution
via the continuous integrals, the magnitude of the self-energy is increased
by about 10 MeV, which is 30\% of the self-energy contribution at this point.
These results for \( \Sigma ^{\rho }_{\pi \pi } \) and \( \Sigma ^{\rho }_{\pi \omega } \)
are used in Eq.\ (\ref{eqn:Chiral_Formula}) to fit the lattice data.

\begin{figure}[t]
{\par\centering \resizebox*{0.9\textwidth}{!}{\rotatebox{90}{\includegraphics{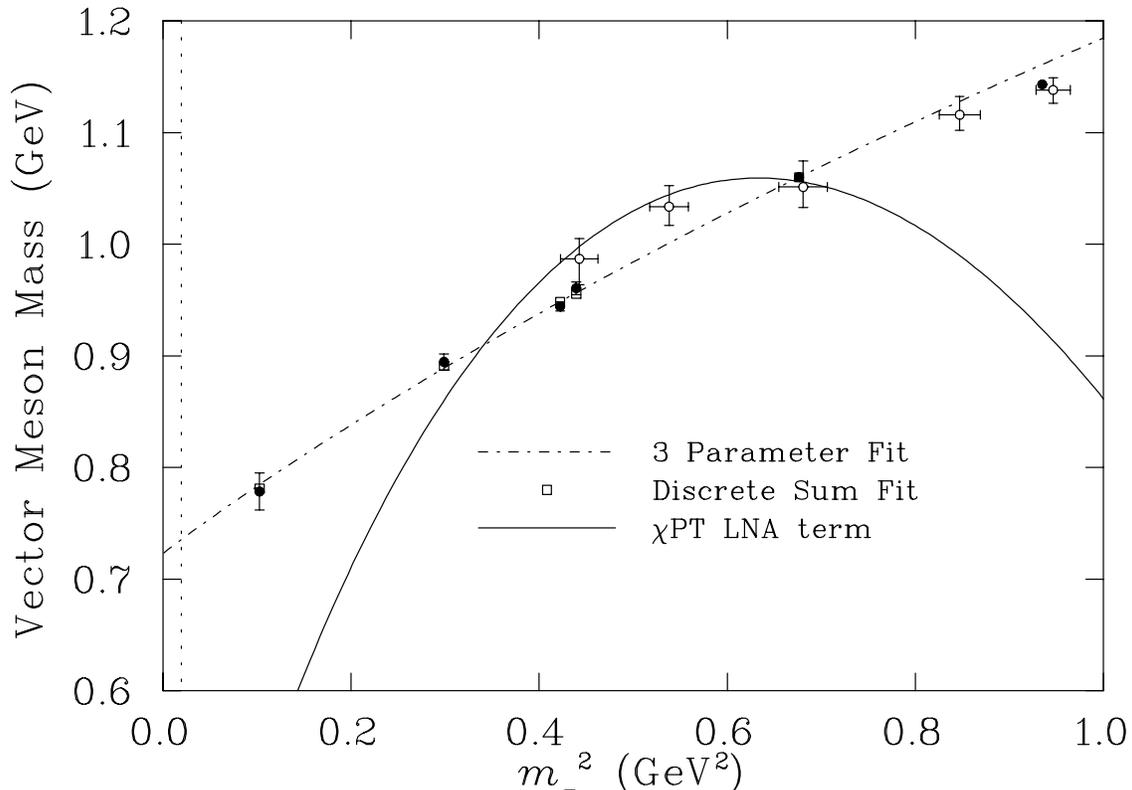}}} \par}

\caption{Vector meson (\protect\( \rho \protect \)) mass from CP-PACS \protect\cite{Aoki:1999ff}
(filled circles) and UKQCD \protect\cite{Allton:1998gi} (open circles) as a
function of \protect\( m_{\pi }^{2}\protect \). The dash-dot curve is the na{\"\i}ve
three parameter fit, Eq.\ (\ref{eqn:3 Parameter Formula}). The open squares
(which are barely distinguishable from the data) represent the fit of Eq.\ (\ref{eqn:Chiral_Formula})
to the data with the self-energy contributions calculated as a discrete sum
of allowed lattice momenta. We have used a dipole form factor, with \protect\( \Lambda _{\pi \omega }=630\protect \)
MeV. The solid curve is Eq.\ (\ref{eqn:3 Parameter Formula}) with the parameter
\protect\( c_{3}\protect \) fixed to the value given by chiral perturbation
theory.\label{fig:Lattice Fit}}
\end{figure}

Recent dynamical fermion lattice QCD results are presented in Fig.\ \ref{fig:Lattice Fit}.
The scale parameters relating the lattice QCD results to physical quantities
have been adjusted \cite{Leinweber:1999ig} by 5\% for the CP-PACS and UKQCD
results. The effect is to increase the \( \rho  \) mass from CP-PACS and decrease
the mass from UKQCD providing better agreement between the two independent simulations.
As the \( \chi ^{2} \) of the following fits is dominated by the CP-PACS data,
we focus on this data set.

\begin{figure}[t]
{\par\centering \resizebox*{0.9\textwidth}{!}{\rotatebox{90}{\includegraphics{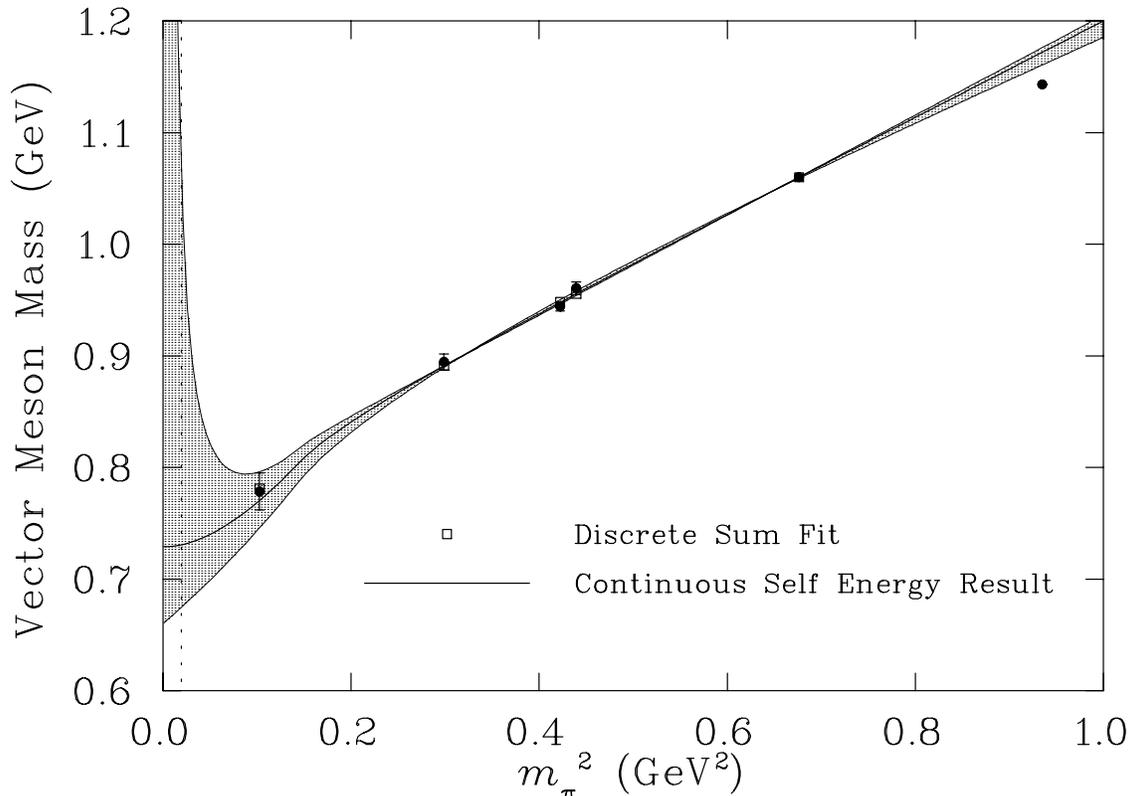}}} \par}

\caption{Analysis of the lattice data for the vector meson (\protect\( \rho \protect \))
mass calculated by CP-PACS as a function of \protect\( m_{\pi }^{2}\protect \).
The squares represent the fit of Eq.\ (\ref{eqn:Chiral_Formula}) to the data
with the self-energy contributions calculated as a discrete sum of allowed lattice
momenta. The solid curve is for continuous (integral) self-energy contributions
to Eq.\ (\ref{eqn:3 Parameter Formula}). We have used a dipole form factor,
with optimal \protect\( \Lambda _{\pi \omega }=630\protect \) MeV. The shaded
area is bounded below by a 1\protect\( \sigma \protect \) error bar. The upper
bound is limited by the constraint \protect\( \Lambda _{\pi \omega }>\mu _{\pi }\protect \)
as discussed in the text.\label{fig:Fit with full error bars}}
\end{figure}

\begin{figure}[t]
{\par\centering \resizebox*{0.9\textwidth}{!}{\rotatebox{90}{\includegraphics{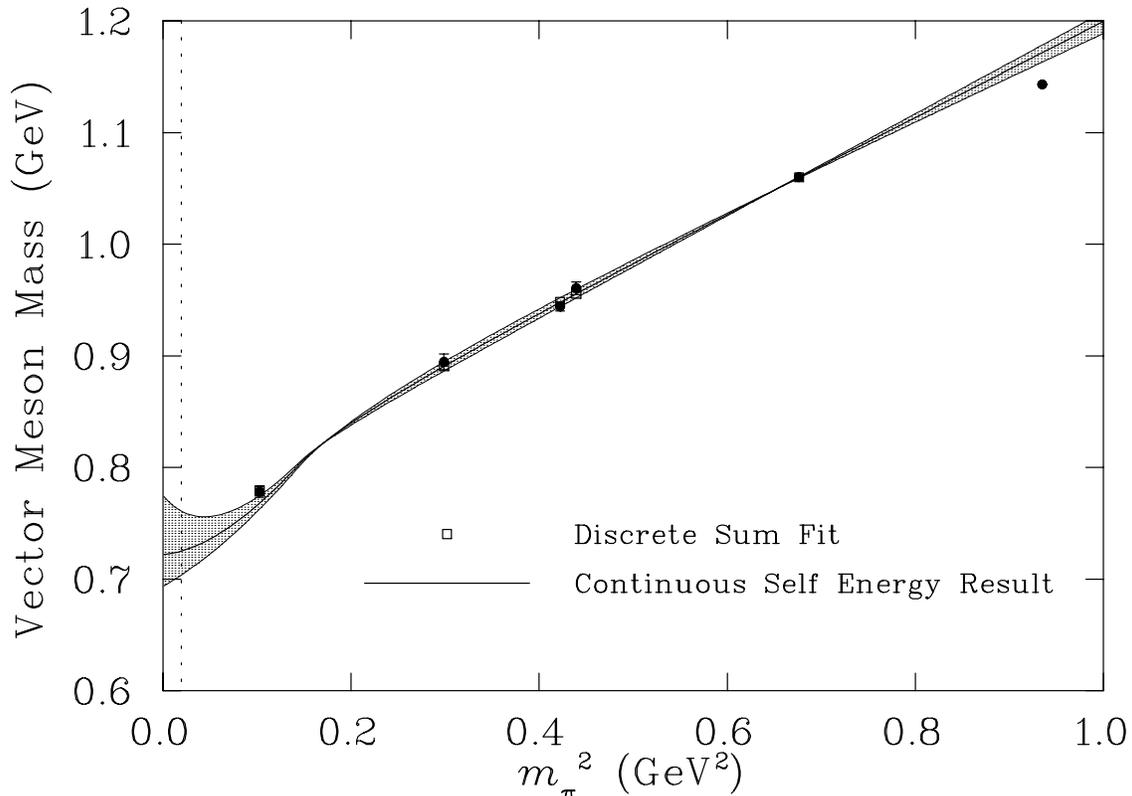}}} \par}

\caption{The graph is as described in Fig.\ \ref{fig:Lattice Fit} except that the error
bar on the lowest data point (\protect\( m_{\pi }^{2}\approx 0.1\protect \)
GeV\protect\( ^{2}\protect \)) has been reduced by a factor of \protect\( \sqrt{10}\protect \).
This equates to an improvement of 10 times in the statistics, which we do not
consider an unreasonable goal for the future. The dipole mass of the best fit
is then \protect\( \Lambda _{\pi \omega }=660\protect \) MeV. The shaded area
is bounded above and below by a 1\protect\( \sigma \protect \) error bar.\label{fig:Fit with root 10 error bars}}
\end{figure}

Our fits using Eq.\ (\ref{eqn:Chiral_Formula}) are based on the lowest five
lattice masses given by CP-PACS. We selected the lowest lying masses because
to move further away from the chiral limit would necessitate additional terms
beyond the first two analytic terms of Eq.\ (\ref{eqn:Chiral_Formula}). \textbf{}The
results of the fit are shown as the open squares in Figs.\ \ref{fig:Lattice Fit},
\ref{fig:Fit with full error bars}, and \ref{fig:Fit with root 10 error bars}.
The parameters of the fit \( c_{0} \), \( c_{2} \), and \( \Lambda _{\pi \omega } \),
are then used in an exact evaluation of Eq.\ (\ref{eqn:Chiral_Formula}) using
the full integrals in Eqs.\ (\ref{eqn:Self Energy pi pi}) and (\ref{eqn:Self Energy pi omega}).
This result is illustrated by the solid lines in Figs.\ \ref{fig:Fit with full error bars}
and \ref{fig:Fit with root 10 error bars}. We note that the value \( \Lambda _{\pi \omega }=630 \)
MeV for the best fit results in a softer form factor than one might expect.
We do not consider this to be of significant concern in the present work because,
as we shall discuss below, the current lattice results at low \( m_{\pi } \)
are not precise enough to constrain the chiral behaviour.

It is interesting to note the similarity of the results to those of Ref. \cite{Leinweber:1994yw}.
There it was found that fitting quenched lattice data with a linear extrapolation,
and improving the extrapolation by adding on the \( \rho \rightarrow \pi \pi  \)
effects, predicted essentially the same physical mass, but that the chiral behaviour
was significantly different.

For comparison we also show a popular three parameter fit, motivated by chiral
perturbation theory:

\begin{equation}
\label{eqn:3 Parameter Formula}
m_{\rho }=c_{0}+c_{2}m_{\pi }^{2}+c_{3}m_{\pi }^{3}\, .
\end{equation}
 
This na{\"\i}ve three parameter fit is illustrated by the dash-dot curve in
Fig.\ \ref{fig:Lattice Fit}. However since the value of \( c_{3} \) in Eq.\
(\ref{eqn:3 Parameter Formula}) is commonly treated as a fitting parameter,
we are not guaranteed that it has the correct value required by Chiral Perturbation
Theory (\( \chi  \)PT). The value for the best fit is found to be \( -0.21 \)
GeV\( ^{-2} \). As outlined above, our expressions for the \( \rho  \) self-energies
have the correct LNA and NLNA coefficients by construction. Indeed, if the coefficient
\( c_{3} \) is constrained to the correct value\footnote{%
In Ref. \cite{Jenkins:1995vb} the \( m_{\pi } \) dependence of the LNA term
to the \( \rho  \) mass is given by \( -\frac{1}{12\pi f^{2}}(\frac{2}{3}g_{2}^{2}+g_{1}^{2})\, m_{\pi }^{3} \).
This would result in a value of the \( m_{\pi }^{3} \) coefficient of \( -1.71 \)
GeV\( ^{-2} \), in excellent agreement with the value used here.
} (\( -g^{2}_{\omega \rho \pi }/48\pi =-1.70 \) GeV\( ^{-2} \)), the best fit
possible by varying \( c_{1} \) and \( c_{2} \) is shown as the solid line
in Fig.\ \ref{fig:Lattice Fit}. As was also found in the case of the nucleon
\cite{Leinweber:1999ig}, the lack of convergence of the formal expansion is
such that it is not sufficient to fix the coefficient of the LNA term in a cubic
fit to that predicted by \( \chi  \)PT, as the resulting form will not fit
the data.

The importance of the accuracy of the lowest mass point cannot be overstated.
We stress that CP-PACS emphasised the preliminary nature of the lowest data
point, because of the relatively low statistics. Nevertheless, in order to prepare
for future more accurate data, we have carried out a standard error analysis
including this point and the results are presented in Fig.\ \ref{fig:Fit with full error bars}.
The lower bound on the shaded area was found by increasing the minimum \( \chi ^{2} \)
per degree of freedom of the fit by 1. We were unable to do this with the upper
bound. The result is actually limited by the physics of the process. In the
case of a dipole form factor this means \( \Lambda _{\pi \omega }>\mu _{\pi } \),
and that is the upper limit we have shown here. 

It is not unreasonable to expect an improvement in the accuracy of the calculated
lattice mass values, and as a Gedanken experiment we have explored the possibility
of a ten-fold increase in the number of gauge configurations at the lowest pion
mass. For the purposes of the simulation we did not change the value of the
data point, but simply reduced the size of the error bar by \( \sqrt{10} \).
 As can be seen in Fig.\ \ref{fig:Fit with root 10 error bars} the improvement
in the predictive power is dramatic. The uncertainty in the physical mass has
been reduced to the 2\% level. Additional improvement in the accuracy of the
extrapolation would result from the availability of additional data in the low
pion mass region. However, it must be noted that the provision of data around
0.2 GeV\( ^{2} \) and higher would probably not assist greatly in the determination
of the dipole mass (\( \Lambda  \)); it is primarily determined by points nearer
the physical region. We present the parameters of these fits in Table \ref{table:Lattice Fit}.

\begin{figure}[t]
{\par\centering \resizebox*{0.9\columnwidth}{!}{\rotatebox{90}{\includegraphics{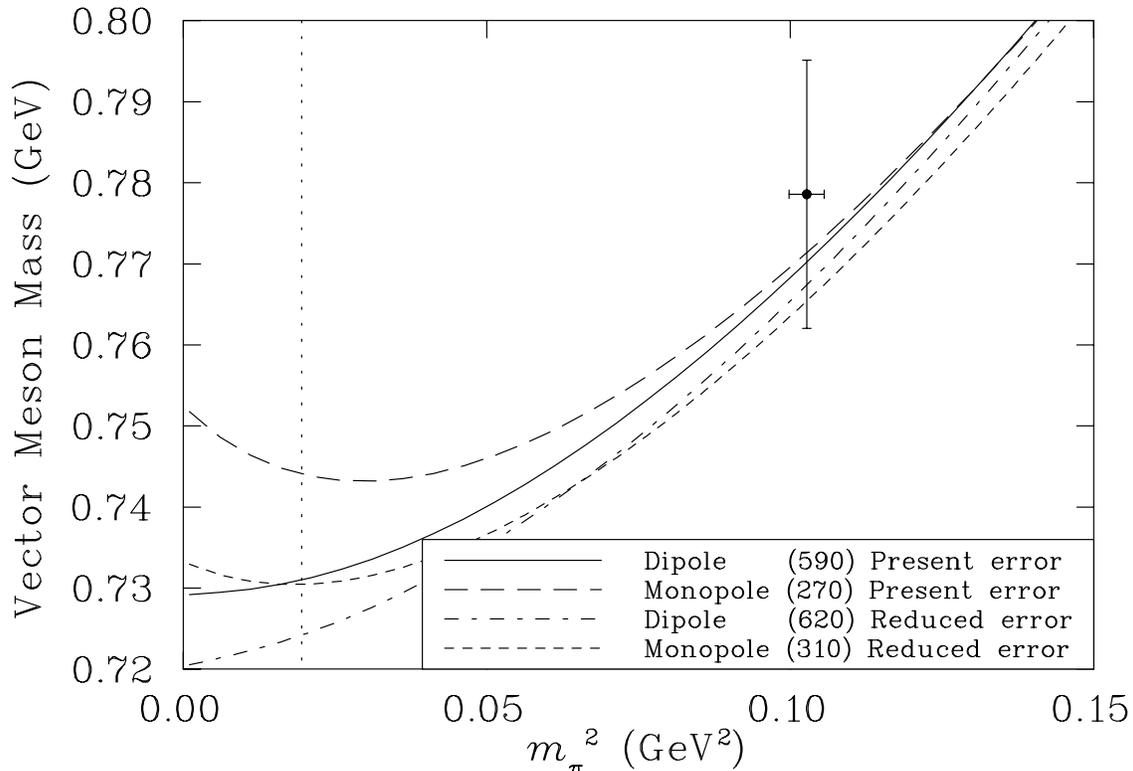}}} \par}

\caption{A magnification of the physical pion mass region of our extrapolation results.
The solid and long dashed lines represent the best fit dipole and monopole results
for a fit with the present accuracy of the lattice QCD results. The dash-dot
and short dashed lines are the dipole and monopole results for a reduction in
the error bar of the lowest lattice data by a factor of \protect\( \sqrt{10}\protect \).
The model dependence of the choice of form factor is \protect\( {\cal O}(2\%)\protect \).\label{fig:Model Comparison}}
\end{figure}

We have examined the model dependence of our work by repeating the above fits
with a monopole form factor. As can be seen in Fig.\ \ref{fig:Model Comparison}
the model dependence is at the level of 15 MeV at the physical pion mass with
current data, and at the few MeV level had the error bar been reduced by a factor
of \( \sqrt{10} \). This reinforces the claim in Ref. \cite{Leinweber:1999ig}
that this extrapolation method is not very sensitive to the form chosen for
the ultra-violet cut-off. 

\section{J - Parameter}

\label{sec:J Parameter}A commonly perceived failure with quenched lattice QCD
calculations of meson masses is the inability to correctly determine the \( J \)-parameter.
This dimensionless parameter was proposed as a quantitative measure, independent
of chiral extrapolations, thus making it an ideal lattice observable \cite{Lacock:1995tq}.
The form of the \( J \)-parameter is:

\begin{eqnarray}
J & = & m_{\rho }\left. \frac{dm_{\rho }}{dm_{\pi }^{2}}\right| _{m_{\rho }/m_{\pi }=1.8}\label{eqn:J-param_Theory} \\
 & \simeq  & m_{K^{*}}\frac{m_{K^{*}}-m_{\rho }}{m^{2}_{K}-m_{\pi }^{2}}\, .\label{eqn:J-param_Expt} 
\end{eqnarray}
By using Eq.\ (\ref{eqn:J-param_Expt}) and the experimentally measured masses
of the \( K \) (495.7 MeV), \( K^{*} \) (892.1 MeV), \( \pi  \) (138.0 MeV)
and \( \rho  \) (770.0 MeV) Lacock and Michael \cite{Lacock:1995tq} determined
\[
J=0.48(2)\, .\]
However previous attempts by the lattice community to reproduce this value have
been around 20\% too small. In the case of quenched calculations this has been
cited as evidence of a quenching error (see, for example the review in \cite{Yoshie:1997hs}).
It was noted by Lee and Leinweber \cite{Lee:1997bq} that the inclusion of the
self-energy of the \( \rho  \)-meson generated by two-pion intermediate states
(excluded in the quenched calculations) acts to increase the \( J \)-parameter.

\begin{figure}[t]
{\par\centering \resizebox*{0.9\textwidth}{!}{\rotatebox{90}{\includegraphics{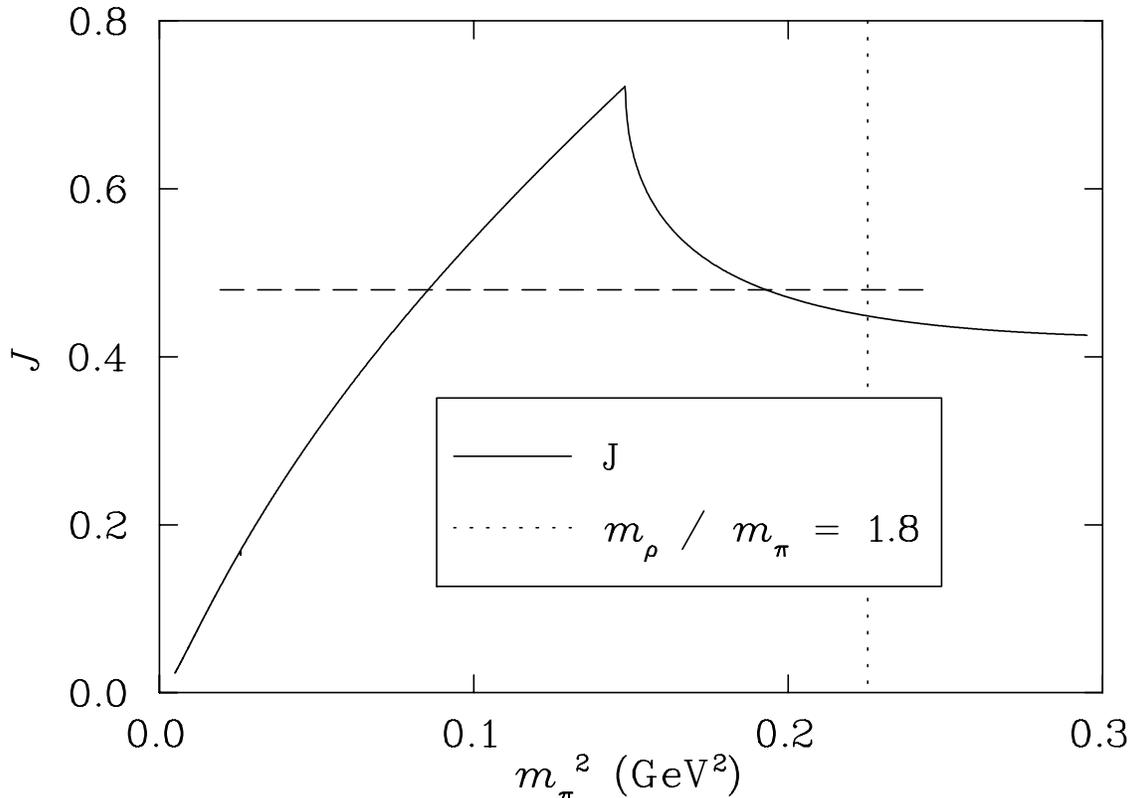}}} \par}

\caption{The solid curve is a plot of the value of the \protect\( J\protect \)-parameter
as a function of \protect\( m_{\pi }^{2}\protect \) obtained from Eq.\ (\ref{eqn:J-param_Theory})
and the best fit to the lattice results given by Eq.\ (\ref{eqn:Chiral_Formula}).
The vertical dotted line shows the point at which the \protect\( J\protect \)-parameter
is evaluated (\protect\( m_{\rho }/m_{\pi }=1.8\protect \)). The horizontal
line displays the experimental value (0.48) plotted between the physical values
of \protect\( m_{\pi }^{2}\protect \) and \protect\( m_{K}^{2}\protect \).\label{fig:J_vs_mpi2}}
\end{figure}
In Fig.\ \ref{fig:J_vs_mpi2} we present the value of the \( J \) parameter
obtained from Eq.\ (\ref{eqn:J-param_Theory}) and our best fit to the lattice
results using Eq.\ (\ref{eqn:Chiral_Formula}). The vertical dotted line indicates
the value of \( m_{\pi }^{2} \) where the \( J \) parameter is to be evaluated,
i.e. \( m_{\rho }/m_{\pi }=1.8 \). The horizontal dashed line, plotted between
the values of the squares of the physical pion and kaon masses, shows the experimental
estimate of the \( J \) parameter from (\ref{eqn:J-param_Expt}). This equation
suggests that the evaluation of \( J \) may be approximated by the slope of
the vector meson mass extrapolation between these points. The cusp shown in
Fig.\ \ref{fig:J_vs_mpi2}, associated with the cut in \( \Sigma _{\pi \pi }^{\rho } \),
suggests otherwise. We stress that while the detailed slope of the curve is
parameter dependent, the presence of the cusp is a model independent consequence
of the two pion cut in the rho spectral function.

As a point of comparison we have also calculated \( J \) using the na{\"\i}ve
cubic chiral extrapolation, Eq.\ (\ref{eqn:3 Parameter Formula}), described
above. The results of our investigations are summarised in Table \ref{table:J-Parameter}. 
\begin{table}[t]
{\centering \begin{tabular}{cccccccc}
Fit Form&
\( c_{0} \)&
\( c_{2} \)&
\( c_{3} \)&
\( \Lambda _{\pi \omega } \)&
\( M_{\rho } \)&
\( J \)&
\( m^{2}_{\pi } \)\\
\hline 
Cubic&
0.723&
0.668&
\( -0.207 \)&
---&
0.735&
\( 0.44\, (8) \)&
\( 0.223\, (7) \)\\
Dipole&
\( 0.776 \)&
\( 0.42732 \)&
---&
\( 0.630 \)&
\( 0.731 \)&
 \( 0.45\, (7) \)&
\( 0.225\, (4) \)\\
\end{tabular}\par}

\caption{Table of fit parameters \protect\( c_{0}\protect \), \protect\( c_{2}\protect \),
\protect\( c_{3}\protect \), \protect\( \Lambda _{\pi \omega }\protect \),
the \protect\( \rho \protect \)-meson mass at \protect\( \mu _{\pi }\protect \),
the value of the \protect\( J\protect \)-parameter, and the pion mass at which
the \protect\( J\protect \) parameter is calculated. All values are in appropriate
powers of GeV. The Cubic fit refers to Eq.\ (\ref{eqn:3 Parameter Formula})
while the Dipole refers to Eq.\ (\ref{eqn:Chiral_Formula}) with a dipole form
factor. We find that the error in the \protect\( J\protect \)-parameter is
halved if the statistics on the lowest point are increased by a factor of 10.\label{table:Lattice Fit}\label{table:J-Parameter}}
\end{table}
The value of the \( J \) parameter is similar for both fits as it is evaluated
at \( m_{\pi }^{2}\sim 0.22 \) GeV\( ^{2} \). The effects introduced into
the extrapolations by chiral physics do not begin playing a large role until
\( m_{\pi }^{2} \) falls below \( 0.2 \) GeV\( ^{2} \). Had the \( J \)
parameter been evaluated at \( m_{\pi }^{2}=0.19 \) GeV\( ^{2} \) or \( 0.09 \)
GeV\( ^{2} \) one would find perfect agreement with the linear ansatz of Eq.\
(\ref{eqn:J-param_Expt}).

\section{Conclusion }

We have explored the quark mass dependence of the \( \rho  \) meson including
the constraints imposed by chiral symmetry. The pionic self-energy diagrams
are unique in that they give rise to the leading (and next-to-leading) non-analytic
behaviour and yield a rapid variation of the meson mass near the chiral limit.
These are the lowest energy states with given quantum numbers that have significant
couplings to the \( \rho  \)-meson. Other meson intermediate states are suppressed
by large mass terms in the denominators of the propagators, and also by smaller
couplings.

We find that the predictions of two-flavour, dynamical-fermion lattice QCD results
are succinctly described by equation (\ref{eqn:Chiral_Formula}) with terms
defined in (\ref{eqn:Self Energy pi pi}) and (\ref{eqn:Self Energy pi omega})
for \( m_{\pi }\leq 800 \) MeV. We have shown that our formula gives model
independent results at the 2\% level for the physical mass of the \( \rho  \)
meson. However, firm conclusions concerning agreement between the extrapolated
lattice results and experiment cannot be made until the systematic errors in
the extraction of the scale of masses can be reduced below the current level
of 10\% and accurate measurements are made at \( m_{\pi }\sim 300 \) MeV or
lower.

We have also calculated the \( J \) parameter by directly evaluating the derivative
of our mass extrapolation formula. We find that the empirical estimate based
on differences of meson masses misses important non-analytic effects in the
derivative of \( m_{\rho } \) with respect to \( m_{\pi }^{2} \), as illustrated
in Fig.\ \ref{fig:J_vs_mpi2}. 

Finally we have investigated the effects of an improvement in the statistics
of the lattice data. Present lattice data is not yet sufficiently precise to
independently constrain the behaviour near the chiral limit. With the best data
available one finds a \( \rho  \)-meson mass of 731 MeV with \( 1\sigma  \)
bounds at 675 and 1062 MeV. One could constrain the bounds by using phenomenological
guidance for the form factors, but we would prefer to wait for better lattice
data. Figure \ref{fig:Fit with root 10 error bars} suggests that the \( \rho  \)-meson
mass could be known to within 5\% in the very near future.

\section*{Acknowledgements}

We would like to thank C. R. Allton, S. R. Sharpe, J. Speth and A. G. Williams
for helpful discussions. We would particularly like to thank A. P. Szczepaniak
for drawing our attention to a correction in the \( \omega \pi  \) self-energy.
This work was supported by the Australian Research Council.

\section*{Addendum}

Since the submission of this manuscript the CP-PACS collaboration has released
a preprint \cite{Khan:2001tx}, with work showing \( J \) as a function of
mass. We note that their analysis does not address the chiral physics studied
here. As a result, their curves will omit the general feature of a cusp in the
\( J \) parameter as discussed in this manuscript. A similar comment applies
to the MILC collaboration preprint \cite{Bernard:2001av}. We look forward to
seeing a similar analysis to that presented here applied to these new simulation
results.

\appendix

\section*{}

In this appendix we present the evaluation of the leading non-analytic terms
of the \( \Sigma _{\pi \omega }^{\rho } \) and \( \Sigma _{\pi \pi }^{\rho } \)
self-energy contributions to the \( \rho  \)-meson mass. By the definition
in Eq.\ (\ref{eqn:Chiral_Formula}) all the non-analytic behaviour is contained
in these two terms.

We note that the form of the self-energy contribution from \( \rho \rightarrow \pi \omega  \)
is the same as that for the process \( \sigma _{NN} \) discussed in Ref.\ \cite{Leinweber:1999ig}.
Using the results found in that paper we can write (for the choice of a sharp
cutoff (\( \theta (\Lambda -k) \)) for the form factor \( u_{\pi \omega } \))
\begin{equation}
\Sigma _{\pi \omega }^{\rho }=-\frac{g_{\omega \rho \pi }\mu _{\rho }}{12\pi ^{2}}\left( m_{\pi }^{3}\arctan \left( \frac{\Lambda }{m_{\pi }}\right) +\frac{\Lambda ^{3}}{3}-\Lambda m_{\pi }^{2}\right) \, .
\end{equation}
The chiral behaviour of this expression is obtained by expanding it in \( m_{\pi } \)
about \( m_{\pi }=0 \) (the chiral limit). We find that in this limit
\begin{equation}
\Sigma _{\pi \omega }^{\rho }=-\frac{g_{\omega \rho \pi }\mu _{\rho }}{12\pi ^{2}}\left( \frac{\Lambda ^{3}}{3}-\Lambda m_{\pi }^{2}+\frac{\pi }{2}m_{\pi }^{3}-\frac{1}{\Lambda }m_{\pi }^{4}+{\cal O}(m_{\pi }^{6})\right) \, ,
\end{equation}
with the leading non-analytic term being of order \( m_{\pi }^{3} \): 
\begin{equation}
\left. \Sigma _{\pi \omega }^{\rho }\right| _{{\rm LNA}}=-\frac{\mu _{\rho }g^{2}_{\omega \rho \pi }}{24\pi }m_{\pi }^{3}
\end{equation}

The \( \rho \rightarrow \pi \pi  \) self-energy contribution is slightly more
complicated. If we again choose a \( \theta  \)-function for the form factor
we can analytically integrate Eq.\ (\ref{eqn:Self Energy pi pi}) giving
\begin{eqnarray}
\Sigma _{\pi \pi }^{\rho } & = & -\frac{f^{2}_{\rho \pi \pi }}{6\pi ^{2}}\frac{1}{2\X }\left( 2\sqrt{m_{\pi }^{2}-\X ^{2}}(m_{\pi }^{2}-\X ^{2})\right. \nonumber \\
 &  & \left\{ \arctan \left( \frac{\Lambda -\X +\sqrt{\Lambda ^{2}+m_{\pi }^{2}}}{\sqrt{m_{\pi }^{2}-\X ^{2}}}\right) -\arctan \left( \frac{\Lambda +\X +\sqrt{\Lambda ^{2}+m_{\pi }^{2}}}{\sqrt{m_{\pi }^{2}-\X ^{2}}}\right) \right. \nonumber \\
 &  & \left. -\arctan \left( \frac{m-\X }{\sqrt{m_{\pi }^{2}-\X ^{2}}}\right) +\arctan \left( \frac{m+\X }{\sqrt{m_{\pi }^{2}-\X ^{2}}}\right) \right\} \nonumber \\
 &  & \left. -(3m_{\pi }^{2}-2\X ^{2})\X \ln \left( \frac{\sqrt{\Lambda ^{2}+m_{\pi }^{2}}+\Lambda }{m_{\pi }}\right) -\Lambda \X \sqrt{\Lambda ^{2}+m_{\pi }^{2}}\right) \, ,
\end{eqnarray}
where \( \Lambda  \) regulates the cut off of the integral. The region in which
we are interested (the chiral limit) has \( m_{\pi }<\X  \). Thus the arguments
of the arctans are complex. We use the relationship
\begin{equation}
\arctan (z)=\frac{i}{2}\ln \left( \frac{1-iz}{1+iz}\right) \, ,
\end{equation}
to rewrite this expression in terms of logarithms with real arguments. Collecting
the logarithms together results in the following expression for the \( \rho \rightarrow \pi \pi  \)
self-energy, for \( m_{\pi }<\X  \):
\begin{eqnarray}
\Sigma _{\pi \pi }^{\rho } & = & -\frac{f^{2}_{\rho \pi \pi }}{6\pi ^{2}}\frac{1}{2\X }\left\{ -\left( \X ^{2}-m_{\pi }^{2}\right) ^{3/2}\right. \nonumber \\
 &  & \ln \left( \frac{m_{\pi }^{2}(m_{\pi }^{2}-\X ^{2})+\Lambda ^{2}(m_{\pi }^{2}-2\X ^{2})-2\Lambda \X \sqrt{(\Lambda ^{2}+m_{\pi }^{2})(\X ^{2}-m_{\pi }^{2})}}{m_{\pi }^{2}(\Lambda ^{2}+m_{\pi }^{2}-\X ^{2})}\right) \nonumber \\
 &  & \left. -\left( 3m_{\pi }^{2}-2\X ^{2}\right) \X \ln \left( \frac{\sqrt{\Lambda ^{2}+m_{\pi }^{2}}+\Lambda }{m_{\pi }}\right) -\Lambda \X \sqrt{\Lambda ^{2}+m_{\pi }^{2}}\right\} \, .
\end{eqnarray}
Looking at just the lowest order, non-analytic, terms in the expansion about
\( m_{\pi }=0 \) we have 
\begin{eqnarray}
\left. \Sigma _{\pi \pi }^{\rho }\right| _{{\rm LNA}} & = & -\frac{f^{2}_{\rho \pi \pi }}{6\pi ^{2}}\frac{1}{2\X }\left( \left( 2\X ^{3}-3\X m_{\pi }^{2}+\frac{3}{4}\frac{m_{\pi }^{4}}{\X }\right) \right. \nonumber \\
 &  & \left. +\left( 3m_{\pi }^{2}-2\X ^{2}\right) \X \right) \ln (m_{\pi })\nonumber \\
 & = & -\frac{f^{2}_{\rho \pi \pi }}{4\pi ^{2}\mu ^{2}_{\rho }}m_{\pi }^{4}\ln (m_{\pi })\, ,
\end{eqnarray}
which is the result given in Eq.\ (\ref{eqn:LNA}).

\bibliographystyle{utcaps}
\bibliography{Me,J-Parameter,Lattice,Rho,General,ChPT,G-E-and-G-M,Extrapolation}

\end{document}